\documentclass[]{spie}  
\usepackage[]{graphicx}

\title{Active X-ray Optics for Generation-X, the Next High Resolution X-ray
  Observatory}

\author{Martin Elvis\footnotemark, 
R.J. Brissenden,
G. Fabbiano,
D.A. Schwartz,
P. Reid,
W. Podgorski,
M. Eisenhower,
M. Juda,
J. Phillips,
L. Cohen,
S. Wolk\\
\medskip
Harvard-Smithsonian Center for Astrophysics, 60 Garden St., Cambridge
MA 02138, USA} 

\footnotetext{Send correspondence to author M.E.; 
e-mail: elvis@cfa.harvard.edu, Telephone: +1 617 495 7442}

\begin{document} 
\maketitle 

\begin{abstract}
X-rays provide one of the few bands through which we can study the
epoch of reionization, when the first galaxies, black holes and stars
were born. To reach the sensitivity required to image these first
discrete objects in the universe needs a major advance in X-ray
optics.  {\em Generation-X} (Gen-X) is currently the only X-ray
astronomy mission concept that addresses this goal. {\em Gen-X} aims
to improve substantially on the {\em Chandra} angular resolution and
to do so with substantially larger effective area. These two goals can
only be met if a mirror technology can be developed that yields high
angular resolution at much lower mass/unit area than the {\em Chandra}
optics, matching that of {\em Constellation-X} ({\em Con-X}).  We
describe an approach to this goal based on {\em active X-ray optics}
that correct the mid-frequency departures from an ideal Wolter optic
{\em on-orbit}. We concentrate on the problems of sensing figure
errors, calculating the corrections required, and applying those
corrections. The time needed to make this in-flight calibration is
reasonable. A laboratory version of these optics has already been
developed by others and is successfully operating at synchrotron light
sources. With only a moderate investment in these optics the goals of
{\em Gen-X} resolution can be realized.
\end{abstract}


\noindent {\bf Keywords:} X-ray optics, active optics, X-ray astronomy

\section{INTRODUCTION}
\label{sect:intro}  

The high angular resolution of the {\em Chandra} X-ray Observatory has
revolutionized X-ray astronomy, and indeed wide areas of astrophysics as a
whole.  At some point though, a larger and higher angular resolution
successor to {\em Chandra} must be built (Fabbiano 1990, Elvis \& Fabbiano
1996). Fortunately X-ray astronomy is nowhere near its physical limits: the
Chandra mirror would have been diffraction limited at $\sim$20~milli-arcsec
(van Speybroeck 2000, private communication), while the thermal line widths
of 1~keV temperature plasmas are of order 100~km~s$^{-1}$, so the 'thermal
limit' to X-ray spectroscopy is $\lambda/\Delta\lambda \sim$6000 (Elvis
2001).

More immediately, Chandra is a small telescope: The median exposure time
for {\em Chandra} observations is a day (85~ksec), and many of the
breakthrough observations needed 1~Msec (Table~1).  This should not be
surprising - the large number of independent spatial and spectral bins {\em
Chandra} offers ($\sim$10$^5$ in ACIS-I) requires large number of photons
to fill them, but {\em Chandra} has an area of only $\sim$800~cm$^{-2}$,
equivalent to a 30~cm diameter optical telescope. As a year contains only
$\sim$20~Msec of net observing time, even at the high observing efficiency
of {\em Chandra}, only a small number of these Megasecond-class programs
can be carried out in a year, and only a limited number in the whole {\em
Chandra} lifetime. A reconaissance of the X-ray sky at high resolution
requires an faster observation rate.

\begin{table}[t]
\caption{Some {\em {\em Chandra}} Breakthroughs}
\begin{tabular}{|lccll|}
\hline
{\em {\em Chandra}} Observations&&&Implication& reference\\
\hline
X-ray background resolved to 10~keV&Ms&$\rightarrow$& 
  accretion luminosity of the Universe&Giacconi et al. 2001\\
Dark Energy measured &&$\rightarrow$& new constraints on 
  new physics& Allen et al. 2004\\
No cooling flows in clusters&&$\rightarrow$&20-year puzzle solved; 
  AGN feedback &David et al., 2001\\
Cooling fronts in clusters&&$\rightarrow$&build up of galaxies, 
  clusters &Vikhlinin et al. 2001 \\
Warm-Hot Intergalactic medium&Ms&$\rightarrow$ &'missing baryons' 
  found& Nicastro et al. 2002 \\
Multi-phase Winds from AGNs &Ms&$\rightarrow$&AGN 'feedback' to galaxy
formation& Kaspi et al. 2002\\
Ultra-Luminous X-ray Sources&& $\rightarrow$& intermediate mass 
  black holes?& Fabbiano et al. 2003\\
Galactic center flares& &$\rightarrow$ & Not so quiet Black Hole 
  & Baganoff et al. 2001\\
Hot ISM abundances&Ms&$\rightarrow$ &SN yields, ecology& 
Fabbiano et al. 2004 \\
Optically dull/X-ray bright galaxies&Ms&$\rightarrow$&hidden 
  active galaxies & Alexander et al. 2003 \\
0.5$c$ wisps in Crab&&$\rightarrow$&particle acceleration to near $c$&
  Hester et al. 2002 \\
\hline
\end{tabular}
Ms: exposure times approaching 1~Msec (2~weeks) needed to make the
discovery. 
\end{table}

In terms of angular resolution, {\em Chandra} often provides just one unique
example of each class of object: e.g. the Crab nebula wisps, the Antennae
galaxy merger system, the 'bullet' cluster 1E~0657-06 (Markevitch et al.,
2004) Only by increasing angular resolution can we expand these samples: a
factor 10 improvement increases the accessible volume 100 times for
Galactic objects (as they lie in a disk), and a 1000 times for
extragalactic objects (and for objects within 100~pc).  At the faintest
fluxes, where the first generation objects will be found, high angular
resolution is required to make firm identifications with counterparts from
JWST, ALMA or other next generation telescopes.

The need then is clear. To give some substance to the science goals we
first describe briefly the first generation of black holes and their
expected properties. Then we outline the {\em Gen-X} Vision Mission study - a
just completed 2-year NASA funded investigation. We then address the
question of active X-ray optics in some detail, to show that this is a
feasible prospect.

\section{Key Science: The First Black Holes} 

The epoch at which the first light was emitted following the Big Bang is
now being pinned down quite closely. The Dark Ages, during which the
temperature of the Big Bang radiation dropped below the point at which it
could keep hydrogen ionized, were clearly over by z=6.4 (an Age of the
Universe of 0.87~Gyr) as we see quasars, Gamma-ray bursts and galaxies
already formed then. As quasars show high metal abundances at z=6, there
must have been more than one generation of supernovae to create these
metals by z=6, pushing the epoch of the first stars backward. In fact, the
3-year WMAP results (Spergel et al. 2006) show an electron scattering
optical depth to the CMB that puts the first ionizing photons at z$\sim$11
(Age = 0.42~Gyr).  Another argument requires the early formation of black
holes: the quasars we know of at z=6 are some of the most luminous objects
in the universe, and the Eddington limit requires that they have masses of
10$^9~M_{\odot}$ or more. But black holes seem highly unlikely to grow at a
rate well above the accretion rate set by the Eddington limit, and so can
only double their masses on the Salpeter timescale,
4.5$\times$10$^7$~yr. The original seed black holes probably were no larger
than a few 100~$M_{\odot}$ (Heger \& Wooseley 2002), so they need $\sim$20
Salpeter times, $\sim$0.9~Gyr to reach their z=6 mass.  At a redshift of
$\sim$10 we should find the first black holes undergoing their most rapid
growth.

How can we observe these objects? Ultraviolet emission is absorbed by
the intervening intergalactic medium (IGM), and redshifting means that
the universe is opaque to all UV and optical wavelengths out at
z=10. The millimeter band is excellent to probe up to high redshifts,
and ALMA will do this. However, at some redshift no molecules will yet
have formed, so this band loses power before the first objects are
reached. Only the radio (SKA), infrared (JWST) and X-ray ({\em Gen-X})
bands can reach back to this epoch, and only X-rays carry strong
atomic features for atoms heavier than hydrogen.

What does it take to see black holes in X-rays during their first growth
spurt? A 10$^3~M_{\odot}$ black hole radiating at the Eddington limit at
z=10 has an X-ray flux of 10$^{-20}$erg~cm$^{-2}$s$^{-1}$ (0.2-2~keV). This
is some 1000 times fainter than the faintest {\em Chandra} deep survey
limit (in 2~Msec, HDF-N, Brandt \& Hasinger, 2005). So an area some 1000
times larger than {\em Chandra}'s is needed to get a detection,
i.e. $\sim$100~m$^2$.  To distinguish these faint background sources from
foreground z$\sim$2-3 galaxies requires a resolution significantly smaller
than their diameters, i.e. $\sim$0.1~arcsec HPD.  These and similar
requirements, derived from a wide range of astrophysics, determined the
basic parameters of {\em Gen-X}

\section{The {\em Generation-X} NASA Vision Mission Study} 
\label{sect:sections}

The {\em Generation-X} mission concept was selected by NASA for study in its
"Vision Mission" program in 2003. The PI of the study was R. Brissenden of
the Smithsonian Astrophysical Observatory (SAO), and there were study team
members from most of the major US institutions involved in X-ray astronomy,
both hardware-oriented and observational (Appendix~A).  The study assumed a
launch date of around 2020.

The Vision Mission study involved several elements:
\begin{enumerate}
\item the development of a comprehensive science case, by a large team of
institutions, that clearly spelled out the flow-down of requirements for
area, spatial resolution, field of view, background, timing and spectral
resolution;
\item optics studies at SAO and GSFC;
\item detector studies at SAO and MIT;
\item mission architecture studies at the GSFC IMDC and the JPL Team-X facilities;
\item student studies of magnetically controlled formation flying (MIT),
optical bench vibration modes (U. Puerto Rico), and of a Kirkpatric-Baez
alternative optic approach (Colorado).
\end{enumerate}

\subsection{Instrument Complement}

A large high resolution X-ray optic can feed photons to a wide variety of
instruments, including those which have not been practical to fly before
due to their photon-hungry nature, e.g. polarimeters (which need
$\sim$10$^{6}$ photons/data point). While {\em Gen-X} may well, and probably
should, carry this class of instrument, there are three devices which {\em Gen-X}
clearly needs to be an astrophysically versatile observatory:
\begin{enumerate}
\item {\em Integral Field Spectrometer:} Envisaged as a microcalorimeter
  array with 2~eV resolution. To cover a 1~arcmin field of view at
  0.1~arcsec requires a 600$\times$600 array;
\item {\em Wide Field Imager:} Envisaged as a silicon based device with
  $\sim$100~eV resolution and a 10~arcmin$\times$10~arcmin field of view
  implying 6~k$\times$6~k arrays, with $>$12~k$\times$$>$12~k being
  desirable to properly sample the PSF;
\item {\em High Resolution Spectrometer:} Envisaged as a reflection grating
  spectrometer reaching $R$=10,000 by means of a ruling density of 11,600
  lines/mm. 
\end{enumerate}

A long focal length X-ray mirror has a large plate scale, 1.4 - 3.6
arcsec~mm$^{-1}$. A 10~armin diameter detector then has an array size of
order 50~cm. A large detector implies a large particle background per
sq. arcsec, so background reduction measures must be carefully considered.

\subsection{Two Mission Architectures}

The two mission architecture studies looked at two distinct ways of
achieving the required area (Figure~1): a '{\em Con-X}-like' approach of 6
multiple identical spacecraft with 8-meter diameter mirrors connected by a
fixed optical bench to a detector system 50 meters away at the focus
(studied by the IMDC), and a 'XEUS-like' approach of a single 20-meter
diameter mirror and a separate free-flying instrument spacecraft at the
other end of the 125-meter focal length (studied by Team-X).

The free-flying alternative has the advantage that only one set of
instruments needed to be built, rather than six. Also, the mirror
is expected to have a longer lifetime than the instruments, so the
free-flying instrument spacecraft architecture gives the option of
replacing the instrument suite with a new instrument spacecraft using
automatic rendezvous, without docking. The free-flying architecture,
however, has the challenge that the mirror-instrument separation must
be actively maintained at all times.  Simple mutual gravitational
attraction will cause the two spacecraft to collide in a short time in
the absence of active control. This is clearly an issue for safe mode
design.

\begin{figure}
\begin{center}
\begin{tabular}{c}
\includegraphics[height=6cm]{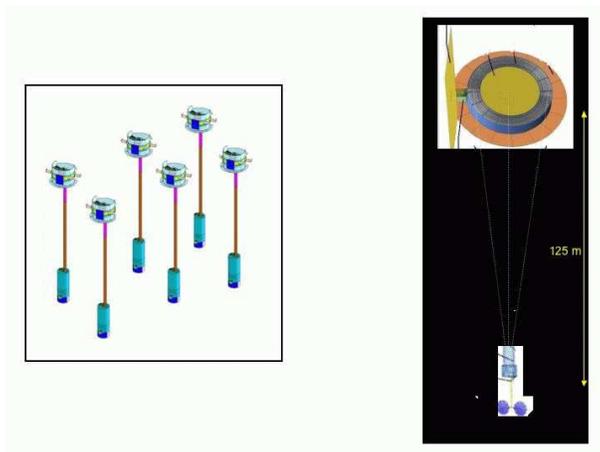}
\end{tabular}
\end{center}
\caption[example] 
{ \label{fig:architectures} The two mission architectures considered
  for {\em Gen-X}: 
{\em left:} 8-meter diameter mirrors in a set of 6 combined
optic-detector spacecraft with 50~m focal length (IMDC);
{\em right:} 20-meter diameter mirror with optics and detector 
spacecraft separated by 125~m (Team-X). } 

\end{figure} 

\subsection{Thermal Control of Optic}

Going in to the study the thermal control of the mirror was a large
concern, as the large mirror area must be unobstructed, and so is free
to radiate to space, yet the mirror needs to be kept isothermal to
$\sim$1~C, and within a range that does not overdeform the 20C figure
formed on the ground beyond the scope of the active correction.
Temperature maintenance could be attained by the use of capillary
pumped loops and both constant and variable heat-pipes to conduct heat
from a thermal collector on the heat/light-shield, solar panel side of
the spacecraft. This has proved to be a robust, low power
solution. The tolerances on the optical bench were acceptable, despite
the 0.1~arcsecond resolution, due to the large plate scale implied by
the long focal lengths.

\subsection{L2 Orbit}

The requirements for a constant thermal environment imposed by the active
optics (see Section 4), the heat-pipe mirror heating scheme and the need
for a low gravity gradient for these long focal lengths, all point to an L2
orbit for {\em Gen-X}. An L2 orbit was recommended by both studies.  

A Delta IV-H launch vehicle is adequate for launching each of the 6
'8-meter' spacecraft and the detector instrument spacecraft of the
`20-meter architecture to L2. The `20-meter' optics spacecraft however
would require dividing the payload into at least two separate launches with
assembly either in LEO or at L2. The `20-meter' approach imposes a launch
window about 21 days wide for the instrument launch with respect to the optic,
in order to match L2 orbits.

The telemetry requirement from L2 was not a major issue, although an
enhancement of current DSN capabilities was assumed. {\em Gen-X} would
generate an average of 400~Gbit of science data per day. Ka band
transmission would require $\sim$1.25 - 1.75 ~hours/day of contact
time.

\subsection{Vision Mission Study}

The Vision Mission study was completed and the report submitted to NASA in
March 2006 and several NASA committees were earlier briefed on the results.

Both studies identified the X-ray optics as the main challenge to
constructing {\em Gen-X}.  The rest of this paper discusses this tall pole.

\section{Active X-ray Optics}

The requirement for high angular resolution combined with low weight
per unit area is tough. The 0.5~arcsec HPD {\em Chandra} mirrors weigh
18500~kg~m$^{-2}$, the {\em Con-X} mirrors weigh 250~kg~m$^{-2}$, while the
13~arcsec HPD XMM-Newton replica mirrors weigh an intermediate
2500~kg~m$^{-2}$. An unsurprising trend of poorer angular resolution
going with lighter weight is apparent. To break this correlation
requires either a dramatic improvement in the manufacture of thin
shell optics, or the approach adopted for the {\em Gen-X} Vision Mission
study - the on-orbit correction of the optic figure by means of active
realignment of small patches of the mirror shells - Active X-ray
Optics.

Active X-ray optics involves adding substantial complication to the X-ray
mirror system. However it comes with several advantages: the ground
calibration of the mirror PSF can be reduced, the launch stability
requirements can be eased, the operating temperature of the mirror need not
be that at which the ground calibration took place (room
temperature). Moreover, compared with the $\sim$10~Hz correction rate for
adaptive optics on ground-based optical telescopes (needed to correct for
the constantly changing atmospheric distortion of the incoming wavefront)
X-ray optics only need adjusting on a months timescale, assuming a benign
constant illumination orbit. The actual rate at which figure adjustment
must be carried out in order not to consume excessive observing time must
be of order days every few months ($\sim$1\%), so the rate is orders of
magnitude slower than ground systems. To maintain almost constant
illumination a baffle/sun-shade system larger than the mirror is required,
and the sun-pointing direction angle must be restricted, e.g. to
$\pm$15$^{\circ}$ of 90$^{\circ}$.

The clear disadvantage is that each 1~sq.m of collecting area at
grazing incidence requires $\sim$100~m$^2$ of reflecting surface which
must be controlled. If corrections need to be made on 10$\times$10~cm
patches this implies 10$^4$ actuators/sq.m collecting area. Depending
on design choices this may rise to 10$^6$ actuators/sq.m (for
1$\times$1~cm patches). The challenges then become: {\em sensing} the
misalignments for all these sensors, {\em calculating} the adjustments to
the figure that are required, and {\em applying} these corrections.

A 0.1~arcsec HPD implies axial figure errors comparable to {\em
Chandra}, but azimuthal figure errors that are substantially tighter.

\subsection{Piezoelectric Actuators}

Ground-based optical systems for active optics use mechanical actuators
that act on the rear of the mirror, perpendicular to the reflecting
surface. While that approach works for normal incidence optics, such a
system would cause significant problems with blockage of the optical path
in the grazing incidence telescopes needed for broad band X-ray
astronomy. Mechanical actuators also need lubricants, which would be hard
to isolate from the reflecting surfaces in the grazing incidence geometry.

\begin{figure}
\begin{center}
\begin{tabular}{c}
\includegraphics[height=7cm]{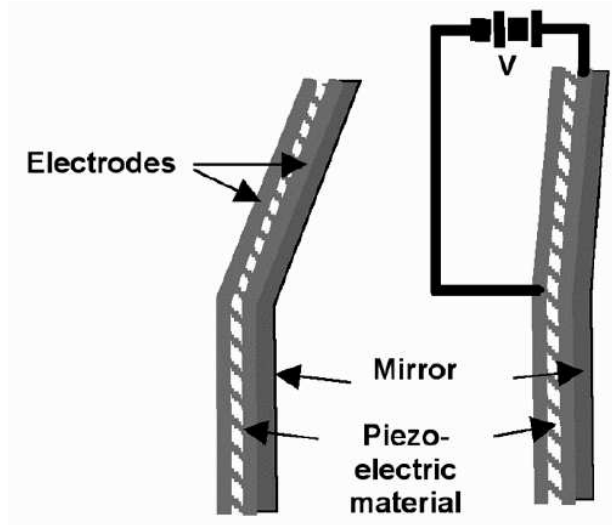}
\end{tabular}
\end{center}
\caption[example] 
{ \label{fig:piezo} Arrangement of piezoelectric actuators on a Wolter I
X-ray optic.}
\end{figure} 

To overcome the optical path blockage problem we explored the possibility
of applying thin piezoelectric actuators to the back side of the X-ray
mirror shells. These act like a bi-metallic strip, bending one way or the
other depending on the applied voltage. A pair of actuators oriented
axially and azimuthally is needed for each patch to be controlled. The
actuators can exert only moderate forces and so are a natural match to thin
shell ($\sim$0.2~mm) optics. Compared to mechanical actuators they have no
hysteresis or backlash, which are clearly undesirable properties.

\subsection{Finite Element Analysis of Piezoelectric actuators}

We assume that the {\em Gen-X} mirror would be constructed from shells
manufactured to a figure PSD tolerance similar to those of the {\em
Con-X} shells (Figure~3). Several other technologies are being
explored that may reach similar tolerances (Hudec et al. 2006, Gubarev
et al. 2006, Friedrich et al., 2006), so we are not presuming a
specific mirror manufacturing technology for {\em Gen-X}.

\begin{figure}
\begin{center}
\begin{tabular}{c}
\includegraphics[height=7cm]{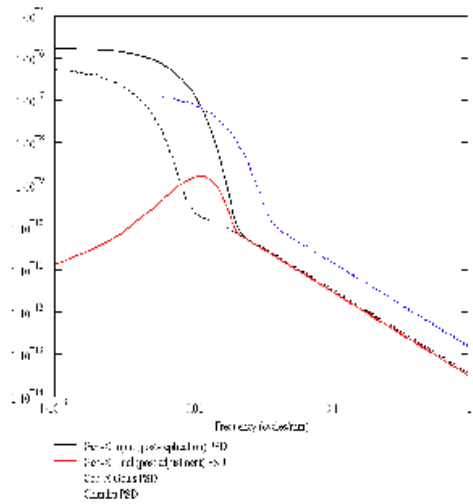}
\end{tabular}
\end{center}
\caption[example] 
{ \label{fig:psd} Power spectral density (power vs. frequency
(mm$^{-1}$) for several mirror figures: the {\em Con-X} goal (light
dashed), {\em Chandra} (heavy dashed), {\em Gen-X} pre-adjustment
(heavy solid), {\em Gen-X} post-adjustment (light solid).}
\end{figure} 

Beginning with {\em Con-X}-class optics, which have demonstrated good
micro-roughness and mid-frequency figure above 0.05~mm$^{-1}$, the
0.1~arcsec HPD requirement for {\em Gen-X} implies correcting
0.4~$\mu$m azimuthal figure errors to 0.004~$\mu$m. We performed a
finite element analysis that shows that this level of correction is
attainable. Assuming 2~cm sized axial correction patches we obtain a
PSD that is well corrected for scales larger than
0.025~mm$^{-1}$. Figure~4 shows how this compares with the {\em
Chandra} PSD. 

The resulting PSF meets the 0.1~arcsec HPD goal at 1~keV (Figure~4) for a
8-meter diameter optic. The longer focal length 20~meter dia. optic has a
less good PSF, due to the smaller graze angles involved. At 0.2~keV
diffraction limits the HPD, while at 6~keV the smaller graze angle again
degrades the PSF. This design is an existence proof. As yet no optimization
has been done on this mirror design to broaden the energy range over which a
0.1~arcsec HPD can be obtained. Nor have we investigated the field of view
over which this tight HPD is obtained, although scaling from
Van~Speybroeck \& Chase (1972), suggests a $\sim$2~arcmin diameter.

\begin{figure}
\begin{center}
\begin{tabular}{c}
\includegraphics[height=10cm]{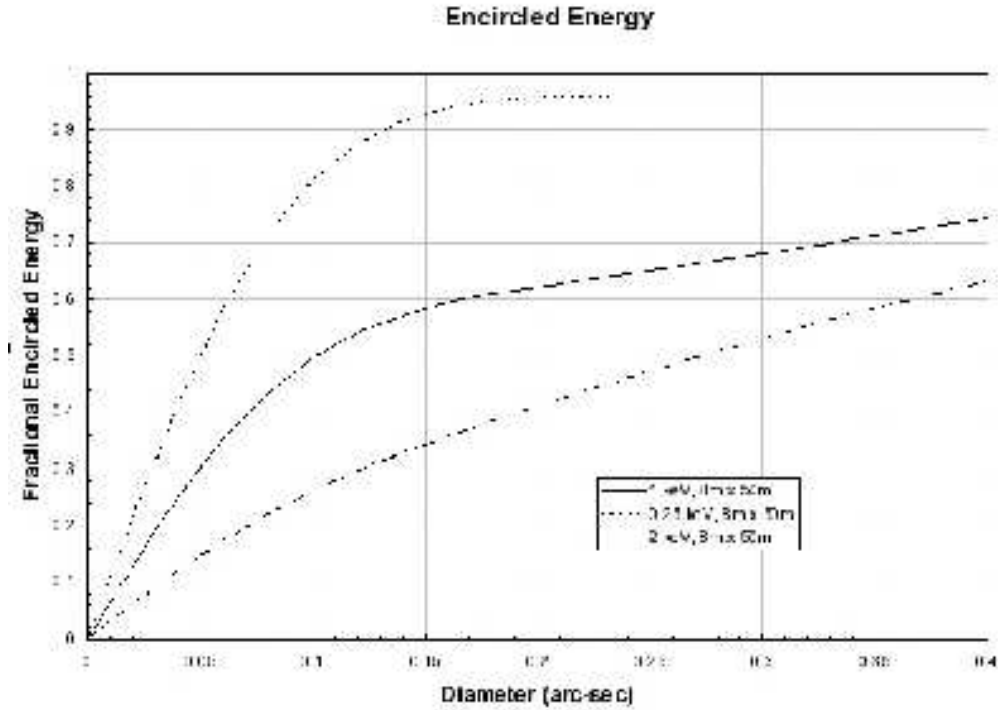}
\end{tabular}
\end{center}
\caption[example] 
{\label{fig:psd} {\em Gen-X} encircled energy vs. radius curve for the 8~m
dia. 50~m focal length design at three energies: 0.25~keV (dotted), 1~keV
(solid) and 2~keV (dot-dashed).}
\end{figure} 

\subsection{Sensing Misalignments \& Calculating Corrections: Out of
  Focus Imaging} 

Each actuator has to have radial, axial and tangential errors measured and
corrected. Yet the on-axis focussed image contains nowhere near enough
information to determine the misalignments of $\sim$10$^6$ mirror patches.

To retrieve this information we explored the possibility of using
out-of-focus images of the converging X-ray beam, thereby separating out
information from every sector of every shell, and resolving each
parabola-hyperbola shell pair axially. This factorizes the problem making
it a weakly-coupled set of equations to solve, rather than a (10$^6$)$^2$
matrix problem. There is no coupling beyond the size of a single mirror
sector.

The inspiration for this approach originated with the {\em Chandra} 'ring focus
test' (Figure~5). (Similar tests have been performed for NeXT and InFOCuS,
Ogasaka et al. 2006). The true ring focus would give a perfect thin ring
for a perfect optic. Instead, for the control of active X-ray optics it is
preferable to have different axial segments of the mirror pairs image to
different locations on the detector in order to resolve them and measure
their misalignments individually.

Placing an imaging detector $\sim$2\% forward of the focus
($\sim$2.5~m for a 125~m focal length), would create separate images
of each shell and of each azimuthal parabola-hyperbola mirror pair
segment.  For example, consider a 20~m diameter, 125~m focal length
mirror employing 10~cm sized active figure correction patches. There
would be $\sim$300 azimuthal sectors in the outermost of the $\sim$
300 mirror shells, and 10 axial segments per reflector (for 1~meter
long reflectors).  Putting a detector 2\% forward of the focus would
give annular images of the mirror shell annuli up to 40~cm in
diameter. These annuli have widths of $\sim$4~cm so each image would
be $\sim$400~$\mu$m thick at this location. These could be imaged into
20 axial elements with a 20~$\mu$m pixel detector, giving the required
resolution of the mirror reflecting surfaces. 

The ``Wide Field Imager'' (\S 3.1) has similar dimensions. In the
`20-meter' architecture the separate detector spacecraft could readily
move this instrument to the correct forward position. In the `8-meter'
single spacecraft option an axial translation stage would be needed.
Whether the out-of-focus detector could double as a wide field
detector at the true mirror focus, or whether a new special purpose
detector that can be inserted into the converging beam as needed is
required, should be the subject of a trade study.

\begin{figure}
\begin{center}
\begin{tabular}{c}
\includegraphics[height=7cm]{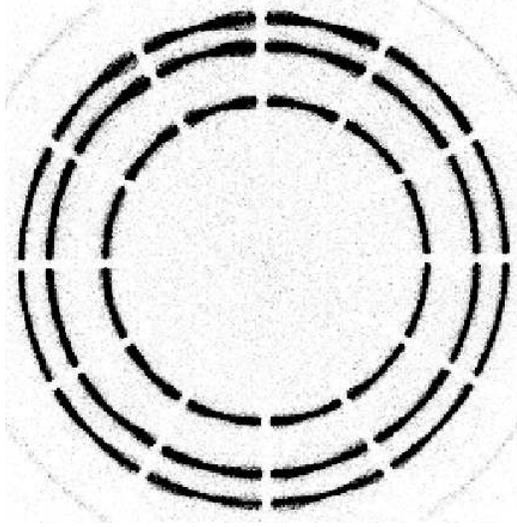}
\end{tabular}
\end{center}
\caption[example] 
{ \label{fig:ringfocus} 
{\em Chandra} (AXAF) ring focus image. The detector was placed forward of the
focus in the converging beam from the four mirror shells (only 3 are shown
here), each of which produces a separate ring that is azimuthally
resolved. For a perfect optic the true ring focus would yield pure circles
for each shell. For an active optics application the detector would
deliberately be placed where a radially extended image would form, imaging
each axial segment of each shell sector separately.}
\end{figure} 

Each square meter of effective area needs 100~m$^{2}$ of reflecting surface
and so $\sim$10$^6$ 1~cm patches. In order to make an accurate correction
let us assume that a 3\% precision is required for each patch and so a
total of 10$^9$ counts m$^{-2}$ of effective area. Is this feasible using
celestial calibration sources? Sco X-1, the brightest persistent source in
the sky, counts at 10$^5$ counts~s$^{-1}$~m$^{-2}$, while the second
brightest source, the Crab Nebula, yields 10$^4$ counts~s$^{-1}$~m$^{-2}$
and some 20 or so other sources count at 10$^3$ counts~s$^{-1}$~m$^{-2}$.
As the number of actuated patches scales with the area, the length of an
alignment observation is independent of the total mission area, and takes
10$^4$s - 10$^6$s/$A_{100}$, where $A_{100}$ is the area of a patch in
units of 100~cm$^2$. This is a reasonable time, and allows for several
cycles of correction during each alignment campaign, unless the patches are
small.

Calculationally, once the problem is factorized by the use of out-of-focus
observations, the problem is not difficult in terms of CPU time. Keck
adjusts 349 actuators at 10~Hz (van Dam et al. 2004). If we require that
the calculation take no longer than the observation then we can operate at
10$^{-4}$Hz - 10$^{-6}$Hz, so allowing 349$\times$10$^{5-6}$ correction
calculations, comparable with the number of actuators, even assuming
no increase in CPU speed.

There is a final degeneracy in the proposed scheme: the pairs of parabola
and hyperbola patches are not separated. This separation could be achieved
by deploying a finite focus calibration source flown as part of the {\em
Gen-X} mission. For such a source illuminating the same parabola patch as a
source at infinity, a different hyperbola patch would be illuminated.
Positioned appropriately, the finite focus source might be able to created
annuli in the gaps between those created by a source at infinity, so that
the two observations could be carried out simultaneously. If the two sets
of images overlap, then a detector with good energy resolution could be
used to separate out an emission line from the finite focus source from the
continuum dominated celestial source. An artificial calibration source
might also provide a higher count rate for small patch mirrors.  This
concept needs further study for feasibility.

\section{Piezoelectric Bi-morph Mirrors at Synchrotrons}

During the course of the Vision Mission study for {\em Gen-X} we searched the
literature for applicable related work. We found that a long term program
of work at synchrotron X-ray light sources has made great progress in
developing piezo-controlled active X-ray optics. The program was motivated
by the need to illuminate small samples without irradiating their
surroundings.

This program, led by R. Signorato (Signorato et al. 2001, 2004), has
developed increasingly complex piezo controlled X-ray mirrors using a
design of pairs of piezos acting oppositely, which removes temperature
effects. As a result they call their designs 'Piezoelectric Bi-morph
Mirrors' (PBMs). These have reached meter-long (by a few cm wide) scales
and operate in the Kirkpatrick-Baez configuration. Hence they need to bend
only axially. Up to 32 actuators are employed, using actuators as small as
2~cm, in a mirror now in use at the Argonne National Labs 'Advanced Photon
Source'.

Signorato et al. find that they can reduce the figure error amplitude by a
factor 15 on the scales they control, from 150~nm to 10nm rms, and further
improvement is expected. The figure correction is stable over days and
months. No anticlastic effect ('saddling') in introduced by the piezos.

To progress from the synchrotron mirrors to ones suitable for X-ray
astronomy requires: adaptation to the 2-D distortions required for a Wolter
geometry; techniques that allow deposition of piezos onto the back of
finished optics (potentially); designs that accomodate the substantial
number of control wires; scaling up of the process from 32 actuators to
10$^4$m$^{-2}$ and the associated questions of speed and cost of
manufacturing.

The independent development of PBMs nonetheless substantially raises
the technical readiness of the piezo-controlled active X-ray optics
needed for X-ray astronomy and {\em Gen-X} in particular, and enable
an accelerated program geared toward demonstration of a flight-ready
optic.

\section{Conclusions}

The future of X-ray astronomy needs a high angular resolution successor to
{\em Chandra}, with larger area. Such a mission would address a huge range of
astrophysics, and in particular would open up one of the few spectral
windows to the study of the very first stars, galaxies, and black holes.

Mission architecture studies show that such a mission is feasible, despite
the long focal lengths (50~m - 125~m) and large diameters (6~m - 20~m). The
tall pole is the development of lightweight high resolution X-ray optics.

Active X-ray optics, controlled with piezoelectric acuators, address the
biggest technical challenge faced by the {\em Generation-X} Vision Mission
concept. Piezos are a good match to the thin shells required by light-weight
optics and to the need not to block the narrow optical path through densely
nested Wolter optics.  A 0.1~arcsec HPD image at 1~keV can be produced with
2~cm sized actuators.  Out-of-focus imaging of the converging beam will
allow figure errors on the individual patches on each sector of each shell
to be measured independently, and celestial sources provide a sufficient
count rate to make the measurements in a reasonable time. The parallization
of the problem afforded by the out-of-focus technique reduces the compute
time to a modest level. Meanwhile, independent work at synchrotrons has
shown that 2~cm sized (1-D) actuators can be constructed, and work well and
reliably, reducing figure errors in a Kirkpatrick-Baez configuration by
large factors.

The {\em Generation-X} Vision Mission study has shown that the
obstacles to creating large high resolution X-ray optics to go well
beyond the capabilities of {\em Chandra} are far smaller than might
have been thought. It has not escaped our attention that once high
resolution, light-weight, X-ray optics are available, then they would
be the mirrors of choice for most, if not all, X-ray astronomy
missions. We believe that the rapid development of active X-ray optics
by means of a quite moderate sized program, is both an urgent and a
reasonable short-term goal for X-ray astronomy.

\acknowledgments 
We thank the other members of the study team and the members of JPL Team-X
and the GSFC IMDC for their major contributions to the {\em Generation-X} Vision
Mission Study. This work was supported in part by NASA Grant NNG04GK28G.



\newpage 
\appendix    
\section{{\em Generation-X} Vision Mission Study Study Team} \label{sect:misc}

\begin{table}[h]
\caption{{\em Generation-X} Vision Mission Study Team}
\label{tab:participants}
\begin{center}       
\begin{tabular}{|l|l|} 
\hline
Team Member& Institution\\
\hline
Roger Brissenden (PI)& SAO\\
Martin Elvis & SAO\\
Giuseppina Fabbiano & SAO\\
Paul Gorenstein & SAO\\
Paul Reid & SAO\\
Dan Schwartz & SAO\\
Harvey Tananbaum & SAO\\
& \\
Richard Mushotzky & GSFC \\
Rob Petre & GSFC \\
Nick White & GSFC \\
William Zhang & GSFC \\
& \\
Martin Weiskopf & MSFC\\
Mark Bautz & MIT \\
Claude Canizares & MIT \\
Enectali Figureoa-Feliciano & MIT \\
David Miller & MIT \\
Mark Schattenburg & MIT \\
& \\
Robert Cameron & Stanford \\
Steve Kahn & Stanford \\
& \\
Niel Brandt & Penn State \\
& \\
Melville Ulmer & Northwestern \\
& \\
Webster Cash & Colorado \\
\hline 
\end{tabular}
\end{center}
\end{table} 

\end{document}